\documentclass[11pt]{article}

\usepackage[total={6.5in,8.75in}, top=2.4cm, left=2.4cm]{geometry}
\usepackage{lineno}
\usepackage{graphicx}

\usepackage{bm}
\usepackage{float}
\usepackage{amsfonts}
\usepackage{verbatim}
\usepackage{soul}
\usepackage{color}
\usepackage{setspace}

\usepackage{amssymb}    

\floatstyle{plain}
\floatname{panel}{Panel}
\newfloat{panel}{h}{txt}

\floatname{panel}{Panel}
\newfloat{panel}{h}{txt}

\title{
Data Augmentation for Hierarchical Capture-recapture Models
}
\author{
{\bf J. Andrew Royle}
\footnote{\emph{email:} aroyle@usgs.gov},
{\bf Sarah J. Converse}\footnote{
\emph{email:} sconverse@usgs.gov 
} and {\bf William
  A. Link}\footnote{
\emph{email:} wlink@usgs.gov 
} \\
 USGS Patuxent
Wildlife Research Center, Laurel, MD 20708
}

\begin{document}

\maketitle

\date

\linenumbers

\begin{flushleft}

{\bf \it Abstract: } Capture-recapture studies are widely used to obtain
information about abundance (population size or density) of animal
populations. A common design is that in which multiple distinct
populations are sampled, and the research objective is modeling
variation in population size $N_{s}; s=1,2,\ldots,S$ among the
populations such as estimating a treatment effect or some other
source of variation related to landscape structure. The problem is
naturally resolved using hierarchical models. We provide a Bayesian
formulation of such models using data augmentation which preserves the
individual encounter histories in the model and, as such, is amenable to
modeling individual effects.  We formulate the model by conditioning
on the total population size among all populations. In this case,
the abundance model can be formulated as a multinomial model
that allocates individuals among sites. MCMC is easily carried out by the introduction of a
categorical individual effect, $g_{i}$, which partitions the total
population size. The prior distribution for the
latent variable $g$ is derived from the model assumed for the
population sizes $N_{s}$.


\end{flushleft}

\section{Introduction}

Capture-recapture models are widely used in ecology and wildlife
management to estimate the size of animal populations (Williams,
Nichols and Conroy 2002). A common situation in many ecological
studies concerns the case where the population is divided into
spatially, temporally referenced or otherwise grouped (henceforth {\it
  stratified}) populations.
This is frequently characteristic of animal
population studies because
spatial replication is often
crucial to the scope of inference. For example, one might establish
$S$ experimental units to investigate the influence of some factor or
treatment on animal abundance, and carry-out a capture-recapture study
on each unit (Dorazio et al. 2005;
Converse and Royle 2012). In addition, in studies of rare or elusive
species that may be difficult to capture, 
 multiple sample units are necessary to obtain a
sufficient sample of individuals in order to effectively estimate
parameters of capture-recapture models (Converse and Royle 2012).

As the number of replicates increases, and especially in the presence
of sparse data or low sample sizes for some of the populations, it
becomes increasingly necessary to aggregate information across
replicates using a model in order to estimate these parameters and to
improve the precision of local population size estimators. Moreover, a
common objective of animal population studies is to develop models
relating local population size to explicit spatial or temporal 
covariates that describe the stratification structure. In practice, this is often done by obtaining local
population-specific estimates of $N$ and then developing additional
models for this collection of estimates, as if they were data (i.e.,
``doing statistics on statistics'' Link 1999).
These two objectives can be addressed
coherently with the use of hierarchical models. In addition, a benefit
of the use
of a formal model-based framework for combining data from multiple populations
is that it explicitly
accommodates variability among spatial sample units so that
valid variance estimates are obtained for estimates of mean population
size across populations, or for predictions on new, unsampled units
or under hypothetical conditions.  Models allow for explicit
characterization of prediction uncertainty, and so there is clear
utility in devising flexible models for providing robust estimates of
variance. 
 
Hierarchical models provide a natural framework for modeling data
from studies on multiple populations and for addressing the important
inference problems of aggregation, prediction and variance
estimation.  In the context of animal population studies based on
capture-recapture methods, we can extend standard capture-recapture
models by including a model component that describes variation in
population size (Royle 2004a; Royle et al. 2004; Dorazio et al. 2005; Royle and Dorazio
2006).  For example, let $N_{s}$ be the size of population
$s$ where the $s=1,2\ldots,S$ populations are organized spatially in
some manner. Regarding the $\{ N_{s}\}$ as latent variables, we might assume
$N_{s} \sim Poisson(\lambda_{s})$ and then focus attention on modeling
the parameters $\lambda_{s}$. Evaluating factors that affect
$\lambda_{s}$ might be the main focus of many studies, but such
hierarchical models also allow us to obtain estimates of the latent
variables $N_{s}$, i.e., the size of specific populations.

We provide general formulations of hierarchical models of abundance
from capture-recapture data by constructing specific classes of prior
distributions for $N_{s}$.  
We provide a framework
for Bayesian analysis that permits analysis of models in which the
dimension parameter space is itself unknown. This arises in classes of
models in which the population size, $N$, is an unknown parameter, and
there are individual-level effects such as heterogeneity or individual
covariates. This variable-dimension parameter space problem causes
considerable difficulty in the analysis of such models using classical
methods of MCMC (Durban and Elston 2005; Royle et al. 2007a; Schofield
and Barker 2011).  Motivated by a need for general Bayesian
treatment of the problem we consider analysis by data augmentation
(Royle et al. 2007a; Royle and Dorazio 2012) which provides a
framework for Bayesian analysis using a {\it fixed} dimension data set
constructed in a specific manner.  One problem with analysis of models
using data augmentation is, because $N$ is not an explicit parameter
in the model under the reparameterization induced by data
augmentation, it is not clear how to model variation in $N$ among
populations, i.e., using hierarchical models, and this has been
suggested as a limitation of data augmentation (Schofield and Barker
2011).  In this paper we resolve this problem by providing a more
general formulation of data augmentation for stratified populations,
in which interest is in modeling variation in $N$ among populations.

To adapt data augmentation for hierarchical models involving
stratified populations, we first develop a multinomial
parameterization of stratified models that conditions on the ``total''
population size. This yields a multinomial formulation of the model
that is suitable for Bayesian analysis by data augmentation.  We also 
consider a  general form of the model using a Dirichlet compound
multinomial model for the population size variables.
Data augmentation for such models is easily implemented in a number of
popular MCMC packages such as WinBUGS (Gilks et al. 1994) and JAGS
(Plummer et al. 2009). As such the inference framework is widely
accessible, e.g., to ecologists, and not just to statisticians trained
in the development of MCMC algorithms.

\section{Data Augmentation}
\label{sec.da}

Data augmentation provides a general approach for analyzing models in
which the multinomial sample size, $N$, is an unknown quantity
(Royle et al.  2007a).  When $N$ is unknown, the dimension of the
parameter space is itself a variable and this requires that
specialized MCMC algorithms be used (e.g., reversible jump MCMC) to
analyze such models (e.g., Durban and Elston 2005; Schofield and
Barker 2008).  Data augmentation effectively reparameterizes models in
which $N$ is an unknown parameter into a model in which $N$ is removed as an
explicit parameter, by marginalizing over a $Bin(M,\psi)$ prior
distribution for $N$, where $M>> N$ is fixed and $\psi$ is the
unknown parameter to estimate. 
The resulting model for the
augmented data can be expressed as a zero-inflated binomial mixture
model and standard Gibbs
sampling methods can be applied to the analysis of the model (Royle
and Dorazio 2012). In
addition, the model reformulated by data augmentation can be expressed
directly in the BUGS language and therefore implemented in popular
software packages such as WinBUGS, OpenBUGS or JAGS.
Data augmentation is especially useful for analysis of models with
individual effects, either in the form of covariates or individual
level random effects because the model preserves an individual
encounter history formulation of the model that is conditional on the
individual effects.

To introduce data augmentation in the relevant context, suppose that a
population of size $N$ is sampled producing encounter data $y_{i}$ on
$n\le N$ individuals.  Without loss of generality we suppose here a
simple observation model in which $y_{i} \sim
\mbox{Binom}(K,p)$ for each individual in the population where $K$
is the number of samples of the population made under the assumption of
closure to movement and population dynamics.  This model is usually
referred to as ``model $M_0$'' in the capture-recapture literature. The
basic inference problem is to estimate  $N$ which is unknown.

Data augmentation is motivated by a specific prior choice for $N$.
For the case of a single population, we assume that $N \sim
\mbox{Binom}(M, \psi)$ for some fixed $M$ and $\psi \sim
\mbox{Unif}(0,1)$ which implies that the marginal prior of $N$ is
$\mbox{Unif}(0,M)$.  In this sense, data augmentation provides a reasonable
specification of a non-informative prior for $N$.  As a conceptual
matter, this prior specification implies a pseudo-population of size
$M$, and a segregation of members of that population into two
sub-sets: (1) ``real'' individuals, which occur with probability
$\psi$, and (2) pseudo-individuals which are not members of the
population of size $N$, and occur with probability
$1-\psi$. This form of the model that arises under data augmentation is convenient for implementation of models with
individual-level effects because we can parameterize the model in
terms of a collection of binary latent variables, say $z_{i}$, which
are Bernoulli trials with parameter $\psi$ such that if $z_{i}=1$,
then the observations $y_{i}$ are generated according to the binomial
observation model and if $z_{i}=0$ then the observations are fixed
zeros with probability 1.  As such, under data augmentation, the model
is a zero-inflated form of the known-$N$ model. 
 In the parameterization of the model based on data
augmentation, the parameter $\psi$ takes the place of $N$ although $N$
can easily be estimated directly (see Royle, Dorazio and Link 2007a) or
obtained as a prediction made from the MLEs under the model for the
augmented model.
As a technical matter, the formulation of data augmentation for the
class of models considered here and by Royle et al. (2007a) is
consistent with ``parameter expansion'' of Liu and Wu (1999) in the
sense that, in addition to data augmentation (Tanner and Wong 1987) an
additional parameter is introduced to accommodate the augmented data
(Royle and Dorazio 2012).

What we do in the remainder of this paper is develop extensions of
data augmentation for models that apply to a
population stratified in some fashion (e.g., spatially or
temporally).  For the case of $S$ 
strata having population sizes ${\bf N} =
(N_{1},N_{2},\ldots,N_{S})$, the goal is to develop models describing
variation in the $N_{s}$ variables and provide an analysis framework
for these models based on data augmentation.  We emphasize that the
$N_{s}$ are latent variables in the context of capture-recapture
models.

\section{Multinomial Abundance Models for Capture-Recapture}

We suppose these $S$ populations are sampled
using some capture-recapture method producing sample sizes $n_{s}$ and
encounter history ${\bf y}_{i}$ for individual $i=1,2,\ldots,
\sum_{s=1}^{S} n_{s}$. 
As the specific
variants of capture-recapture observation models are not the main
focus of this paper,  we address only the most basic model in which
encounter histories are reduced to the scalar encounter frequency,
$y_{i} \sim \mbox{Binom}(K,p)$ where necessary.
Although we note that treatment of, for example, sampling occasion effects in capture probabilities can easily be accommodated. 
Let $g_{i}$ be a covariate (integer-valued, $1, \ldots,
S$) indicating the population membership of individual $i$. This
covariate is {\it observed} for the sample of captured individuals but
not for individuals that are not captured.

A key idea that we develop shortly is that specific priors for $g_{i}$
are consistent with certain reasonable priors for the population size
variables $N_{s}$. Then, the data from all populations can essentially
be pooled, and analyzed as data from a single population with the
appropriate model on $g_{i}$.  The partially observed population
membership variable $g_{i}$ is a {\it categorical} type of individual
covariate (Huggins 1989; Alho 1990; Royle 2009).  In the analysis of
this model using data augmentation that we develop subsequently, the
assumption of a model for the collection of abundance variables
$N_{s}$ {\it implies} a specific model for the individual covariate
$g_{i}$.  That is, data augmentation for stratified populations is
equivalent to an ``individual covariate'' model with a specific
distribution for the individual covariate. Shortly we will consider
two models for $g_{i}$ derived from Poisson and negative binomial
models for $N_{s}$.

To illustrate this data structure, we suppose that a population
comprised of 4 sub-populations is sampled $K=5$ times. Then a
plausible data set has the following structure:
\begin{verbatim}
      individual (i) : 1  2  3  4  5  6  7  8  9 10  
      frequency (y)  : 1  1  3  1  1  2  2  4  1  1
      group (g)      : 1  1  1  2  3  3  3  3  4  4
\end{verbatim}
This data set indicates three individuals were captured in subpopulation 1
(captured 1, 1, and 3 times), a single individual was captured in
population 2, four individuals were captured in population 3, and two
individuals were captured in subpopulation 4.

We suppose some distribution for the subpopulation size parameters,
\[
N_s \sim f(N;\lambda_s)
\]
for $s=1,2,\ldots,S$.  We consider specific forms of $f(N)$ below. To
model variation in $N_{s}$ as a function of some covariate, $x(s)$,
thought to affect variation in the population sizes, we consider
models of the form:
\[
 \log( \lambda_{s} ) = \beta_0 + \beta_{1}*x(s).
\]
The general strategy we adopt is to formulate the joint prior
distribution for the $N_{s}$ parameters by conditioning on the total,
say $N_{T} = \sum_{s} N_{s}$ which is the super-population of all
individuals alive in the $S$ populations. We consider multinomial
prior distributions for the subpopulation sizes:
\begin{equation}
{\bf N}| N_{T} \sim \mbox{Multinom}( {\bm \pi} |N_{T})
\label{eq.multinomial}
\end{equation}
 with specific forms of the cell probabilities ${\bm \pi}_{s}$
dictated by the choice of $f(N)$. This multinomial model forms the
basis of our data augmentation scheme for multiple populations.

\subsection{Poisson case}

We begin with the Poisson case because this model is the 
model commonly used in ecology for modeling count data, and so it
is natural  as a model for the unknown population size parameters.
We assume
\begin{equation}
 N_{s} \sim \mbox{Poisson}(\lambda_{s})
\label{eq.poisson1}
\end{equation}
with
\begin{equation}
\log( \lambda_{s} ) = \beta_{0} + \beta_{1} x(s)
\label{eq.poisson2}
\end{equation}
where $x(s)$ is some measured attribute for population $s$. Under this
Poisson model, by conditioning on the total population size over all
$S$ populations, the $N_{s}$ variables have a multinomial distribution:
\begin{equation}
{\bf N} = (N_{1},\ldots,N_{S}) | \{ N_{T} =
\sum_{s} N_{s} \} \sim \mbox{Multinom}( {\bm \pi} | N_{T}).
\label{eq.mn.N}
\end{equation}
with 
multinomial probabilities $\pi_{s} = \lambda_{s}/\sum_{s} \lambda_{s}$.

To devise a data augmentation scheme for this model of population
size, we  embed the multinomial for $\{ N_{s} \}$ into a
multinomial of the same dimension but with larger, fixed sample size. 
Specifically, we introduce a latent super-population variable $G_{s}$
which we assume has the desired Poisson distribution but with scaled mean:
 $G_{s} \sim \mbox{Poisson}(A \lambda_{s})$ where $A>>1$ where $A$ 
exists (can be chosen) to ensure that $G_{s}$ is arbitrarily larger than $N_{s}$.
Conditional on the total
super-population size $M = \sum_{s} G_{s}$, then  ${\bf G}$ has a
multinomial distribution: 
\begin{equation}
{\bf G}|M \sim \mbox{Multinom}(M;  {\bm \pi} ) 
\label{eq.mn1}
\end{equation}
where $\pi_{s} = \lambda_{s}/\sum_{s} \lambda_{s}$ which are the same
probabilities as for the target multinomial for ${\bf N}$. 
This multinomial model for the super-population sizes $G_{s}$ is
equivalent to the following:
\[
 g_{i} \sim \mbox{Categorical}({\bm \pi} )
\]
for $g_{i}; i=1,2,\ldots, M$.
Given {\bf G} or, equivalently, $g_{i}$, we specify a model for $\{ N_{s}\}$ that differentiates
between ``real'' and ``pseudo-'' individuals by a Bernoulli sampling
model:
\[
 N_{s} \sim \mbox{Binom}(G_{s} , \psi)
\]
where $\psi \sim \mbox{Unif}(0,1)$. Bernoulli sampling preserves the
marginal Poisson assumption (Takemura 1999). That is, $N_{s}$ is
Poisson, unconditional on $G_{s}$ and, also, 
conditional on $N_{T} = \sum_{s} N_{s}$,
${\bf N}$ has a multinomial with probabilities ${\bm \pi}$ and
index $N_{T}$.  Note also that $N_{T} \sim \mbox{Binom}(M, \phi)$
which is consistent with  data augmentation applied to total
population size $N_{T}$. This binomial sampling model can be
represented, equivalently, by the set of Bernoulli variables:
\[
 z_{i} \sim \mbox{Bern}(\psi)
\]
for $i=1,2,\ldots,M$.

The multinomial construction makes it clear that $\psi$ is confounded
with $\exp(\beta_{0})$. By constructing the model
conditional on the total, we lose information about the intercept
$\beta_{0}$, but this is recovered in the data augmentation parameter
$\psi$.  One of these parameters has to be fixed. We can set $\beta_0
= 0$ or else we can fix $\psi$.  The constraint can be specified by
noting that, under the binomial data augmentation model
$E[N_{T}] = \psi M$ and,
under the Poisson model, $E[N_{T}] = \sum_{s} \exp(\beta_{0} + \beta_{1}
x(s))$ and so we can set
\[
 \psi = \frac{1}{M} \sum_{s} \exp(\beta_{0} + \beta_{1} x(s)).
\]
The equivalence of $\psi$ and $\beta_{0}$ can be thought of in terms of pooling data
from the different sub-populations. In a model with {\it no} covariates, 
we could pool 
all of the data and estimate a single parameter $\psi$ or $\beta_0$ but not
both. In this sense, 
 pooling data from multiple spatial samples is justifiable (in terms of
sufficiency arguments) under a Poisson assumption on local abundance
(which was noted by Royle 2004b; Royle and Dorazio 2008, sec. 5.5.1).

\subsection{Implementation}
\label{sec.implementation}

By introducing the latent $G_{s}$ structure, and the Bernoulli
sampling
of $N_{s}$,
the model is equivalently represented by 
the 
latent variable pair $(g_{i},z_{i})$ where $g_{i}$ is categorical
with prior probabilities $\pi_{s}$ and $z_{i} \sim Bern(\psi)$.  
In particular, 
the multinomial assumption for the latent variables $G_{s}$
is formulated in terms of  ``group membership'' for  each individual
in the super-population of size $M$ according to:
\[
 g_{i} \sim \mbox{Categorical}\left( {\bm \pi} \right)
\]
with ${\bm \pi} = (\pi_{1}, \ldots, \pi_{S})$ and $\pi_{s} =  \lambda_{s}/(\sum_{s}
     \lambda_{s})$.
Note that aggregating these $M$
categorical variables yields a set of multinomial variables
consistent with Eq. \ref{eq.mn1}. That is, define 
$G_{1} = \sum_{i=1}^{M} I(g_{i} = 1)$, 
$G_{2} = \sum_{i=1}^{M} I(g_{i} = 2)$, etc., where $I()$ is the
indicator function. 
The binomial sampling from the super-population, 
$N_{T} \sim \mbox{Binom}(M, \psi)$
can be described at the level of the individual also, 
 by introducing the binary
variables $z_{1},\ldots,z_{M}$ such that
\[
 z_{i} \sim \mbox{Bern}(\psi)
\]
where $\psi$ is constrained as noted in the previous section. 
We implement this individual-level formulation of the model in BUGS in
Panel \ref{panel.wbcode}.

A second implementation of the model is suggested by working from
Eq. (\ref{eq.mn.N}) -- we can marginalize $N_{T}$
over the prior  $N_{T} \sim \mbox{Binom}(M, \phi)$ to see that 
the  $(S+1) \times 1$ vector 
$(N_{1},\ldots,N_{S},N_{S+1})$ has, conditional on $M$, 
a multinomial distribution 
with cell probabilities
$\pi_{s}^{+} = \pi_{s} \psi$ for $s=1,2,\ldots,S$  and 
 $\pi_{S+1}^{+} = (1-\psi)$ for the last cell which
 corresponds to individuals of the super-population that are not
 members of any of the $S$ populations that were subject to sampling. 
Thus,
\[
{\bf N}|M \sim \mbox{Multinom}({\bm \pi}^{+}).
\]
where the superscript $+$ here indicates that ${\bm \pi}$ is a larger
version of ${\bm \pi}$ from \ref{eq.mn1}.
In this case, 
\begin{equation}
g_{i}  \sim \mbox{Categorical}( {\bm \pi}^{+} ) \mbox{ for
  $i=1,\ldots,M$}  \label{eq.parm1c}
\end{equation}

\subsection{Negative Binomial Model}
\label{sec.nb}


It is natural to consider abundance models that exhibit
over-dispersion relative to the Poisson model. Here we generalize the
model formulation of the previous section for that purposes.  We
introduce a collection of $iid$ latent super-population size variables
$G_{s}$ which have a negative binomial distribution and we apply the
binomial sampling model of 
data augmentation to these super-population sizes.
 Then,
conditioning on the total super-population size $M = \sum_{s}
G_{s}$, this produces the Dirichlet compound multinomial (DCM) distribution
for the vector $(G_{1},\ldots,G_{S})$
(Takemura 1999).
Equivalently, the DCM arises by placing a
Dirichlet(${\bm \alpha}$) prior on the multinomial probability vector
${\bm \pi}$.

We take the approach of a Gamma-poisson mixture 
(following Takemura (1999)) in which  $G_{s}$ has a Poisson
distribution conditional on $\lambda \eta_{s}$ where $\eta_{s} \sim
\mbox{Gamma}(\alpha,\beta)$ and the $G_{s}$ are mutually independent. 
Marginalizing $\eta_{s}$ from the conditional Poisson model produces
the negative binomial having pmf
\[
 \Pr(G_{s}) = \frac{\Gamma(\alpha +  G_{s})}{\Gamma(\alpha) G_{s}!}
\frac{ (\lambda \beta)^{G_{s}}}{  (1+ \lambda \beta)^{G_{s} + \alpha}
}
\]
so  the mean $\lambda$ is preserved as a parameter in this
formulation: the expected value of $G_{s}$ is $E[G_{s}] = \alpha \beta
\lambda$. Therefore we require a constraint among the parameters
$\alpha$ and $\beta$.
In our analysis below we set $\beta = 1$.  In fact, the total
$M=\sum_{s} G_{s}$ is the
sufficient statistic for $\beta$ in the Gamma-Poisson
mixture\footnote{An ecological example of a DCM application
is Link and Sauer (1997) who
use the DCM for modeling nuisance variation in the BBS. In that
  case, 
$\eta_{s}$ is the ``observer effect''.}
Therefore, 
the two free parameters to
estimate are $\alpha$ and $\lambda$. Preserving $\lambda$ in the model
is convenient for modeling covariates -- which we can do in the usual way:
\[
\log( \lambda_{s} ) = \beta_{0} + \beta_{1} x(s)
\]
for some covariate $x(s)$.

To apply data augmentation to this model, we relate the 
population size variables $N_{s}$ to $G_{s}$ by the binomial
sampling model:
\[
N_{s} \sim \mbox{Binom}(G_{s}, \psi)
\]
which allows us to implement the model using $iid$ Bernoulli trials
$z_{i} \sim \mbox{Bern}(\psi)$ and an individual covariate $g_{i} \sim
\mbox{Categorical}({\bm \pi})$ as before. In this case, we construct
the cell probabilities $\pi_{s}$ according to
\[
\pi_{s} =  \frac{ \eta_{s} \lambda_{s} }{ \sum_{s} \eta_{s}
  \lambda_{s} }
\]
where $\eta_{s}$ is the gamma noise term, and $\lambda_{s}$ represents
the fixed effects.
Implementation of this model in the various BUGS engines poses no
special difficulty, see
Panel \ref{panel.wbcode2}.

Marginally $G_{s}$ is negative binomial and under this Bernoulli
sampling model the $N_{s}$ are also marginally independent negative
binomial models (Takemura 1999, Appendix A) or, when conditioned on
the total $N_{T} = \sum_{s} N_{s}$ is Dirichlet compound multinomial (DCM). 

\section{Capture-recapture with individual effects}

The individual-level formulations of the model have  utility
in the analysis of general capture-recapture models because it allows
for the development of individual-level models for the encounter
probability independent of the models describing variation in $N$
among populations.  Models that affect abundance determine the form
of the cell probabilities ${\bm \pi}$ whereas models of encounter
probability are parameterized directly at the level of individual
encounters. For example, if $y_{i}$ are the individual encounter
frequencies, then the most basic model is:
\[
 y_{i} \sim \mbox{Binom}(K; p z_{i})
\]
so that if $z_{i} = 0$ then $y=0$ with probability 1. In general we
can express the model so that $p$ varies by individual, sample
occasion, stratum or according to some specific covariate. Therefore, a
more general formulation conditions on the stratum membership variable
$g$:
$y_{i,k} | g_{i}  \sim \mbox{Binom}(K; p_{i,k}(g_{i})z_{i})$ with additional individual-level model structure imposed on
$p_{i,k}(g_{i})$.
The simplicity of the observation model is retained regardless of the
complexity of models for $p$, and this yields considerable flexibility
in model development and, importantly, separates the development of
such models (for encounter probability) from the development of models
describing variation in $N$ among subpopulations. 

We provide an implementation of the
capture-recapture model for stratified populations in 
Panel \ref{panel.wbcode}.  This is for the Poisson model in which 
\[
 \log(\lambda_{s}) = \beta_0 + \beta_1 x(s).
\]
The full specification of the model shown in Panel \ref{panel.wbcode}
for the Poisson model.
The BUGS model specification for the model in which $N_{s}$ have the
joint Dirichlet compound multinomial distribution is shown in Panel
\ref{panel.wbcode2}. Note in the 2nd case, we demonstrate the
alternative formulation in which $\psi$ is a free
parameter with $\beta_{0}$ constrained to 0.

The Poisson model shown in Panel \ref{panel.wbcode} 
 is summarized by the following elements:
\begin{eqnarray*}
\mbox{{\it Observation model}: } &  & \\ 
y_{i}|z_{i}& \sim& \mbox{Binom}(z_{i}p, K) \\
\mbox{{\it Process model}: } & & \\
z_{i} & \sim & \mbox{Bern}(\psi) \\
g_{i} & \sim & \mbox{Categorical}( {\bm \pi}) \\
\end{eqnarray*}
where
$\pi_{s} = \lambda_{s}/\sum \lambda_{s}$. The prior distributions are: 
\begin{eqnarray*}
p    & \sim & \mbox{Unif}(0,1)\\
\beta_{0},\beta_{1} & \sim & \mbox{Normal}(0,.1) \\
\psi &  \equiv &  \frac{1}{M} \sum_{s} \exp(\beta_{0} + \beta_{1} x(s)).
\end{eqnarray*}
Note the BUGS language uses a parameterization of the normal
distribution in terms of the $\tau = 1/\sigma^{2}$.

\begin{panel}[htp]   
\renewcommand{\baselinestretch}{1.0}
\centering
\rule[0.15in]{\textwidth}{.03in}
\begin{minipage}{3.5in}
{\small
\begin{verbatim}
model {
# This version constrains psi with 
#   the intercept parameter
  p~ dunif(0,1)
  b0~dnorm(0,.1)
  b1~dnorm(0,.1)
  psi<- sum(lam[])/M

  for(s in 1:S){
    log(lam[s]) <- b0 + b1*x[s]
    gprobs[s]<- lam[s]/sum(lam[1:S])
  }
  for(i in 1:M){
    g[i] ~ dcat(gprobs[])
    z[i] ~ dbern(psi)
   y[i]~ dbin(mu[i],K)
   mu[i] <- z[i]*p
  }
  N <- sum(z[1:M]) 
}
\end{verbatim}
}
\end{minipage}
\rule[-0.15in]{\textwidth}{.03in}
\caption{BUGS model specification for a capture-recapture model with
  constant encounter probability and Poisson subpopulation sizes,
  $N_{s}$, with mean depending on a single covariate \mbox{\tt x[s]}. 
}
\label{panel.wbcode}
\end{panel}

\begin{panel}[htp]   
\renewcommand{\baselinestretch}{1.0}
\centering
\rule[0.15in]{\textwidth}{.03in}
\begin{minipage}{3.5in}
{\small
\begin{verbatim}
model {
  p~ dunif(0,1)
 b0~ dnorm(0,.1)
 b1~ dnorm(0,.1)
 a~dunif(0,1000)
 b<-1              # set to 1 -- trades-off with "psi"
 psi~ dunif(0,1)   # instead, psi is estimated here

  for(s in 1:S){
    eta[s] ~ dgamma(a,b)
    lam[s]<- b0 + b1*x[s]
    alpha[s]<- eta[s]*lam[s]
    gprobs[s]<- alpha[s]/sum(alpha[])
  }
  for(i in 1:M){
    g[i] ~ dcat(gprobs[])
    z[i] ~ dbern(psi)
   mu[i] <- z[i]*p
    y[i] ~ dbin(mu[i],K)
  }
  N <- sum(z[1:M]) 
}
\end{verbatim}
}
\end{minipage}
\rule[-0.15in]{\textwidth}{.03in}
\caption{BUGS model specification for a capture-recapture model with
  constant encounter probability and Dirichlet compound-multinomial
  population sizes.
}
\label{panel.wbcode2}
\end{panel}

\section{Application: Effect of Forest Management on Dear Mouse Populations}

Here we consider a typical problem which motivates the need for
hierarchical models of the type considered in this paper. The data
come from Converse et al. (2006), and are based on a mark-recapture
study of small mammals examining the ecological impacts of forest
thinning and prescribed burning treatments on 13 study areas across
the US (P. Weatherspoon and J. McIver, United States Department of
Agriculture Forest Service, unpublished report). Here we focus on data
collected at the Southwest Plateau Study Area, near Flagstaff,
Arizona.  The Southwest Plateau Study Area was composed of 3 study
sites (blocks), each of which was in turn composed of 4 experimental
units. Live-trapping grids were used to catch small mammals in each of
these 12 experimental units in each year from 2000 to 2003. Twelve
experimental units by 4 years produced 48 groups.
Thus, the sub-populations (strata) in our analysis are indexed by an aggregate
``space $\times$ time'' index $k=1,2,\ldots, 48$.
Before data were collected in 2003, thinning
treatments were applied to 2 experimental units at each study site.
The deer mouse (Peromyscus maniculatus) combined with a small number of individuals of the closely related brush mouse (P boylii) create the dataset we analyze here (489 individuals; range = 0-45 per experimental unit per year). 
Live-trapping occurred in the late summer on
a grid with 25-m spacing (50-m spacing in the first year). Traps were
baited with grain and checked mornings and afternoons for 5 days to
yield 10 trapping occasions (in 2000, 6 of the experimental units were
trapped for only 9 occasions). Captured individuals received 2
uniquely-numbered ear tags.

The main objective is to evaluate the effect of treatment (forest
thinning) on population size. Importantly, the response variable
$N_{s}$ is a latent variable and cannot be measured directly, and thus
hierarchical models are necessary to integrate data from the 48
replicates and to explicitly incorporate the biological hypothesis
(effect of thinning) into the model for abundance.

In our analysis, following Converse et al. (2006), we fitted a model
that contains the treatment effect (1 parameter), and fixed block (2
parameters), and year (3 parameters) effects.  We parameterize the two
block effects and year effects using dummy variables \mbox{\tt B1} and
\mbox{\tt B2} and \mbox{\tt YR1}, \mbox{\tt YR2} and \mbox{\tt YR3}.
The population size model is
specified by:
\[
\log(\lambda_{s}) = \beta_0 + \beta_1 \mbox{\tt trt}_{s} 
+ \beta_{2} \mbox{\tt B1}_{s}
+ \beta_{3} \mbox{\tt B2}_{s}
+ \beta_{4} \mbox{\tt YR1}_{s}
+ \beta_{5} \mbox{\tt YR2}_{s}
+ \beta_{6} \mbox{\tt YR3}_{s}
\]
and, for 
the group membership variables $g_{i}$,
\[
 g_{i} \sim \mbox{Categorical}({\bm \pi})
\]
where
\[
\pi_{s}          = \frac{ \lambda_{s} }{ \sum_{s} \lambda_{s}}.
\]
For the data augmentation variables we have:
\[
 z_{i} \sim \mbox{Bern}(\psi)
\]
and the observation model is
\[
 y_{ik} \sim \mbox{Bern}( p_{ik} z_{ik} ).
\]
We consider a detection probability model that contains 
a behavioral response according to:
\[
 \mbox{logit}(p_{ik}) = \alpha_0 + \alpha_1 y_{i,k-1} 
\]
Here, $\alpha_1$ affects the increase or decrease 
in encounter probability for previously
captured individuals.
This model was fitted in WinBUGS using a 
modification of that shown in Panel \ref{panel.wbcode}.

\subsection{Goodness-of-Fit}

We evaluated the goodness-of-fit of the model described previously
using a Bayesian p-value (Gelman et al. 1996). The statistic was based
on the adequacy of the model at explaining the variation in
population-specific sample sizes of observed individuals, $n_{s}$.
For each posterior sample, $m$, we computed a fit statistic based on
Pearson residuals according to
\[
X_{obs}^{m} =   \sum_{s=1}^{48}
\frac{ 
[
n_{s} -  n_{T} \pi_{s}^{m} 
]^2 
}
{  n_{T} \pi_{s}^{m} }
\]
where 
 $n_{s}$ is the number of individuals encountered in
population $s$,
$n_{T} = \sum_{s} n_{s}$,
 and $\pi_{s}$ is the probability that an individual
in the super-population of size $N_{T}$ belongs to population $s$.
Note that $\pi_{s}^{m}$ depends on model parameters and therefore for
each iteration of the MCMC algorithm, its value changes. 
The fit statistic computed from the observed data is compared to that
obtained by samples from the 
posterior predictive distribution of the data.
That is, for each iteration $m$ of the MCMC algorithm, we obtain 
a draw of $n_{s}^{m}$, a prediction of the number of individuals
captured for population $s$ and $n_{T}^{m}$ is the total sample size
of the simulated populations.
The fit statistic is then computed using the simulated data:
\[
X_{sim}^{m} =   \sum_{s=1}^{48}
\frac{ 
 [
    n_{s}^m -  n_{T}^{m} \pi_{s}^m 
  ]^2 
}
{  
  n_{T}^{m} \pi_{s}^m 
}
\]

\begin{figure}[htp]
\centering
\includegraphics[height=4in,width=7in]{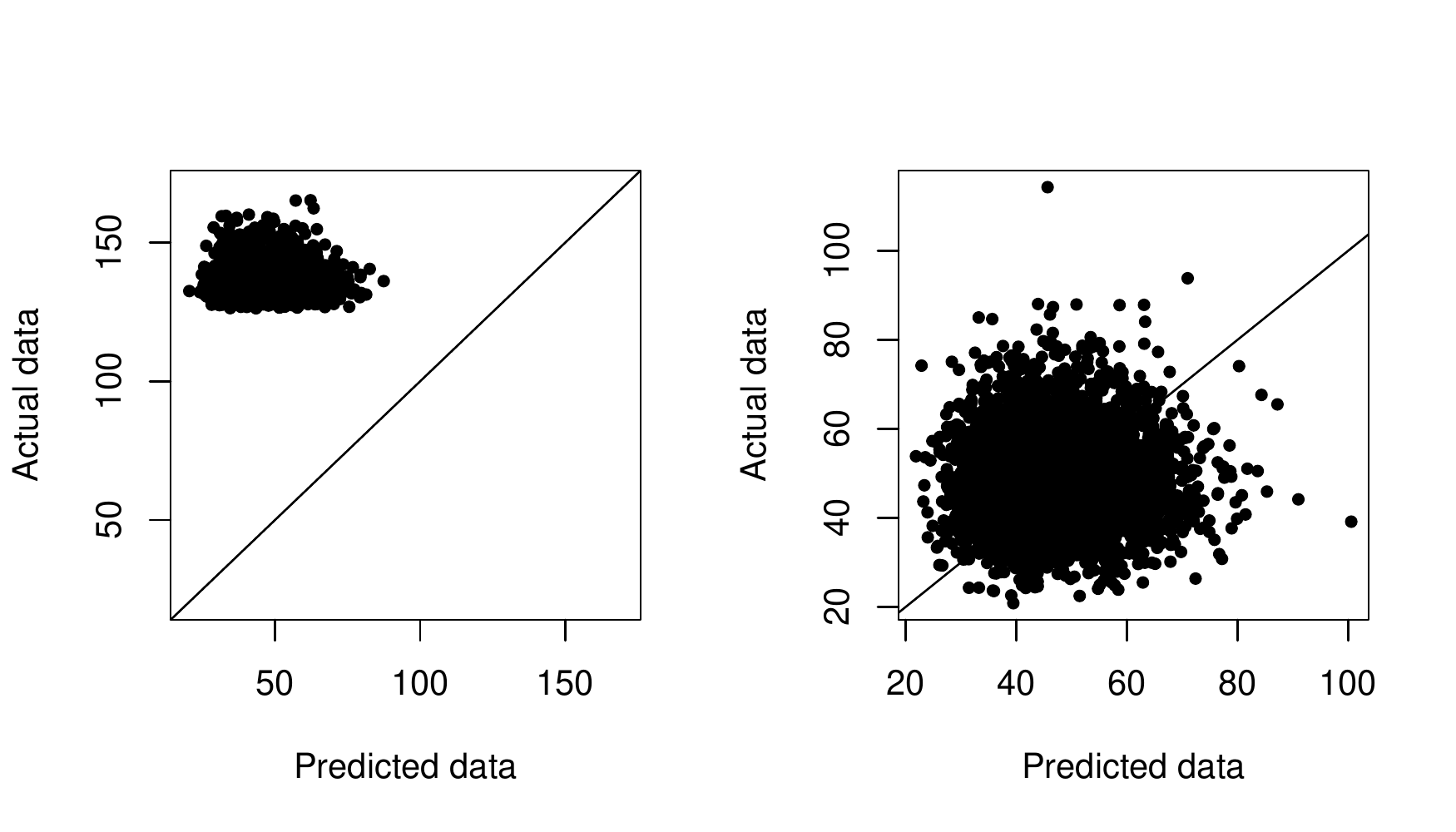}
\caption{
Scatter of fit statistic for observed data vs. posterior simulated
data sets under Poisson (left panel) and Dirichlet compound
multinomial models (right panel).
}
\label{fig.fig1}
\end{figure}

\begin{figure}[htp]
\centering
\includegraphics[height=5in,width=5in]{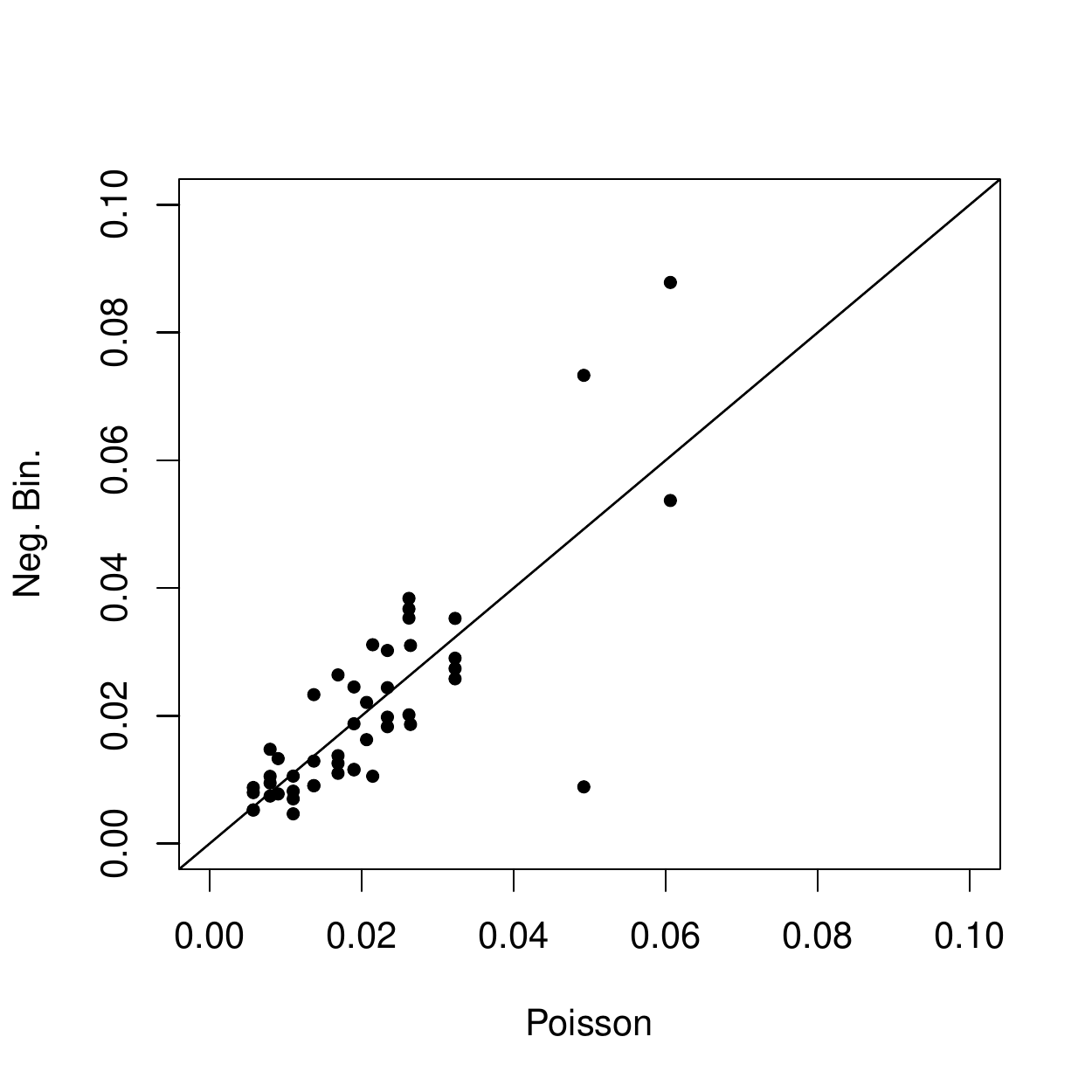}
\caption{
Posterior means of subpopulation probabilities, $\pi_{s}$, under
Multinomial (Poisson) and Dirichlet compound multinomial (negative
binomial) models. 
}
\label{fig.fig2}
\end{figure}

We believe that this fit statistic should 
emphasize the fit (or lack therefore) of the spatial component of the
model which is our main interest here. The scatter of 
$X_{obs}$ vs $X_{sim}$ for the Poisson model fitted to the Peromyscus
data is shown in the
left panel of Fig. \ref{fig.fig1}. The p-value is $0.000$ for that
case indicating a clear lack of fit. Conversely, the fit under the
Dirichlet compound-multinomial base model (with treatment, block and
year effects) (right panel) yields a p-value of $0.558$ indicating
that this model provides an adequate fit to the data.  We assessed
convergence by looking at the Brooks-Gelman-Rubin ``R-hat'' statistics
(Gelman and Rubin 1992; Brooks and Gelman 1997) for the structural
parameters of the model. For both Poisson and DCM models, those were
all less than 1.1 (see Table \ref{tab.microtus})
 and so we think the MCMC output is
adequate for characterizing features of the posterior distribution.
Other posterior summaries are provided in Table \ref{tab.microtus}.
 To summarize the salient results, the 
treatment effect is somewhat strong under both models, and comparable
in magnitude, but less precise under the DCM model as we might
expect, because under this model we expect more background variation
in population sizes $N_{s}$.  
How different is the estimated distribution of the population among
the 48 groups?
We plot the estimated population membership probabilities, $\pi_{s}$,
under the two models in 
Fig. \ref{fig.fig2}.  In general, there is less shrinkage to the mean
under the DCM model -- 
we see one site where the Poisson model 
over-predicts substantially and 2 moderate under-predictions but
otherwise most of the population proportions are
similar.
Both models indicate a very strong {\it positive} behavioral response,
indicating strong trap-happiness.
Total population size
is slightly higher under the DCM model (posterior mean: roughly 710 vs. 695) and the
posterior is only slightly more diffuse under the DCM model.

\begin{table}
\centering
\caption{
Posterior summaries of 
multinomial and Dirichlet compound multinomial
hierarchical capture-recapture models fitted to 48 populations of {\it
  Microtus}. The two models correspond to the multinomial abundance
model which arises under a Poisson assumption, and the Dirichlet
compound multinomial model which arises under a negative binomial
assumption for abundance.
}
\begin{tabular}{rrrrrrr} 
\hline \hline
\multicolumn{7}{c}{Multinomial Model} \\
parameter &      mean &    sd &    $2.5\%$ &       $50\%$ & $97.5\%$ &  Rhat  \\ \hline
$\beta_0$     &     1.700 & 0.148  &  1.340  &    1.704   &   1.970 &1.011     \\
$\beta_1$     &     0.835 & 0.161  &  0.522  &    0.834   &   1.166 &1.001  \\
block[2]  &     0.872 & 0.132  &  0.632  &    0.865   &   1.147 &1.001  \\
block[3]  &     1.080 & 0.130  &  0.844  &    1.077   &   1.342 &1.001  \\
year[2]   &    -0.327 & 0.150  & -0.626  &   -0.326   &  -0.026 &1.007  \\
year[3]   &     0.324 & 0.132  &  0.070  &    0.326   &   0.586 &1.014  \\
year[4]   &     0.118 & 0.166  & -0.222  &    0.122   &   0.427 &1.006  \\
$\alpha_0$ (intercept)   &    -2.038 & 0.133  & -2.264  &   -2.044   & -1.783 &1.019 \\
$\alpha_1$ (trap happy)   &     0.495 & 0.142  &  0.214  &    0.500   & 0.755 &1.029 \\
$\psi$       &     0.637 & 0.045  &  0.558  &    0.636   &   0.728 &1.014   \\
N         &   695.064 & 47.019 & 616.000 &   693.000  &  788.000&1.016   \\
\hline \hline
\multicolumn{7}{c}{Dirichlet compound multinomial model} \\
parameter &      mean &    sd &    $2.5\%$ &       $50\%$ & $97.5\%$ &  Rhat  \\ \hline
$\beta_0$     &     0.439&  0.433 &     0.151 &   0.379 &     1.385 &1.020 \\  
$\beta_1$     &     0.781&  0.338 &     0.559 &   0.790 &     1.432 &1.010  \\  
block[2]  &     0.832&  0.226 &     0.680 &   0.835 &     1.273 &1.008  \\  
block[3]  &     1.012&  0.225 &     0.869 &   1.018 &     1.446 &1.012  \\  
year[2]   &    -0.340&  0.256 &    -0.514 &  -0.340 &     0.161 &1.006  \\  
year[3]   &     0.225&  0.249 &     0.067 &   0.224 &     0.710 &1.006  \\  
year[4]   &     0.114&  0.296 &    -0.087 &   0.115 &     0.694 &1.009  \\  
$\alpha_0$ (intercept)    &    -2.078&  0.146 &    -2.175 &  -2.070 &    -1.798 &1.007  \\  
$\alpha_1$ (trap happy)   &     0.540&  0.156 &     0.431 &   0.532 &     0.854 &1.007  \\  
$\psi$       &     0.652&  0.054 &     0.615 &   0.645 &     0.771 &1.004  \\  
N         &   710.786& 56.170 &   671.000 & 703.000 &   839.000 &1.005  \\  
\end{tabular}
\label{tab.microtus}
\end{table}

\section{Discussion} 

Understanding variation in population size, $N$, in structured
populations is a fundamental interest in animal ecology. A common
situation is that in which populations are stratified (spatially,
temporally, etc..).
Questions related to how $N$ responds to treatments or landscape
variation are routine in ecology and natural resource studies.
Unfortunately, $N$ is not observable in practical field situations. As
a result, many methods have been devised for estimating $N$ from
individual level encounter history data (e.g., Williams et al. 2002).
A common approach to studying variation in the size, $N$,  among populations
is to obtain estimates of $N$ using such capture-recapture methods,
and then regard such estimates as ``data'' in a second-stage procedure
(Converse and Royle 2012).
 A more contemporary approach to modeling
variation in $N$ is to use hierarchical models (Royle and Dorazio
2006, 2008; Link and Barker 2009; 
K\'{e}ry and Schaub 2011) in which the observation model is formulated conditional
on $N$, and the model is extended to include an additional component
describing variation in $N$.

In principle, Bayesian analysis of stratified  abundance models is
straightforward without data augmentation, for certain classes
of models in which parameters do not vary at the level of
individual. In such cases, encounter histories can be aggregated into
frequencies for which the joint distribution is a simple multinomial
with index $N_{s}$. Various prior distributions can be specified for
$N_{s}$, and the resulting model analyzed directly using standard
Bayesian or frequentist methods.  One example is given in Royle et
al. (2007b) and similar models can be found in, for examples, Royle et
al. (2004a) using a distance-sampling model, and Dorazio et al. (2005)
for a removal type of sampling protocol. 
Certain classes of multinomial-mixture models can be fitted by
likelihood methods in 
the software package \mbox{\tt unmarked} (Fiske and Chandler 2012)
using the functions
 \mbox{\tt
  multinomPois} and \mbox{\tt gmultmix}
(see Chandler et al. 2011).
 However, direct analysis of
such models by MCMC is more difficult when they contain 
individual-level effects, and that motivates the need for methods
based on data augmentation as developed in our paper. 
Another approach suggested for analyzing such
models is reversible jump MCMC (RJMCMC; e.g., Schofield and Barker
2008, 2011). 
Unlike the data augmentation
approach, RJMCMC retains the collection of population size parameters
$N_{s}$ in the model and relies on a specialized MCMC algorithm to
move about the state-space of $N_{s}$ while reconciling the
dimensionality of the remaining parameters.  Conversely, in analyzing
these models by data augmentation, as we have done in this paper, the
$N_{s}$ parameters are removed from the model by marginalizing over
the next level in the hierarchical model.  In the present case, the
joint prior distribution of ${\bf N}|N_{T}$ is marginalized over
the prior for $N_{T}$. The advantage of removing the $N_{s}$
parameters from the model is that the dimension of the parameter space
is {\it fixed} and thus standard Gibbs Sampling algorithms can be used
to update model parameters. It has been suggested that a deficiency
with the use of data augmentation to analyze capture-recapture models 
is that because $N$ is not
retained as an explicit parameter, alternative priors or models could
not be imposed on population size. Our work here resolves that problem
in some generality.  In particular, considering multinomial or Dirichlet
compound-multinomial models yields great flexibility in modeling
variation in $N_{s}$ among subpopulations using capture-recapture
observations models of arbitrary complexity at the individual level.

A key idea of our models is conditioning on the total population size
among the distinct populations -- the super-population size of
individuals alive in all of the populations combined. 
This is a similar concept to that employed in the
Crosbie-Manly-Schwarz-Arnason (CMSA) formulation (Schwarz and Arnason 1996)
of the Jolly-Seber model which involves a temporal sequence of
populations $N_{t}; t=1,2,\ldots,T$.  However, in that case,
parameterization of the population sizes $N_{t}$ is based on a
Markovian survival/recruitment model as the sub-populations may share
the same individuals, but with a state-variable (alive, or not) which
evolves over time due to recruitment and mortality.  A formulation of
the CMSA model using data augmentation was given in Royle and Dorazio
(2008, ch. 10), and K\'{e}ry and Schaub (2011, ch. 10).
In the context of the CMSA model, the $g_i$ variable is ``period of entry'' into the
population.
Our model seems
relevant to formulating open population models of the CMSA variety. In
particular, with $g_{i}$ being the
period of entry into the population, we could suppose that
\[
 g_{i} \sim \mbox{Categorical}({\bm \pi})
\]
where $\pi_{t} = \lambda_{t}/(\sum_{t} \lambda_{t})$, and therefore we
can model factors that affect recruitment directly on $\lambda_{t}$,
Our analysis therefore suggests that the CMSA formulation of the model
implies a Poisson recruitment model, where the total number of
recruits in year $t$, $R_{t}$, is Poisson with mean
$\lambda_{t}$. Alternatively, the DCM model could be used as a model
for recruitment to accommodate variability across years.

The other key idea of our model development here is that we embed this
``super-population'' in the CMSA sense into yet a larger population --
a super-super-population -- of size $M$, which facilitates the use of
data augmentation (Royle et al. 2007b).  Data augmentation produces an
individual-level formulation of models for stratified populations in
terms of the group membership variable $g_{i}$, which is a categorical
type of individual covariate as in Huggins (1989), Alho (1990), and
Royle (2009) -- $g_{i}$ is observed for each individual in the sample
but not for unobserved individuals.  In particular, the assumption of
a model for $N_{s}$ {\it implies} a specific model for the individual
covariate $g_{i}$.  Therefore, data augmentation for spatially
stratified populations is equivalent to an ``individual covariate'' model
with a specific distribution for the individual covariate.

We provided an illustration of our model formulation to a small-mammal
trapping study wherein the main objective was to evaluate a treatment
effect on population size.  Historically such analyses have been
carried out by using closed population models to obtain estimates of
$N$ for each grid and then treat the estimates as data in a
second-stage regression procedure (e.g., Converse et al. 2006). Our
analysis of these data involved a behavioral response model, which is
almost always essential in small-mammal trapping studies.  Thus our
model was a multi-population version of this important type of closed
population model.  
 Although we considered
a single specific encounter probability model (``model $M_b$'') we
could as well consider any other model containing individual-level
effects, including  alternative behavioral response formulations
(Yang and Chao 2005), individual heterogeneity (Pledger 2000; Dorazio
and Royle 2003) or spatially explicit capture-recapture models
(Borchers and Efford 2008; Royle and Young 2008).

The method is also relevant to other problems in which abundance is
naturally stratified.  A common situation in wildlife surveys involves
sampling groups of individuals (Royle 2008; Royle 2009), such as
flocks of birds, herds of ungulates, or family groups of marine
mammals.  In such cases, encounters of individuals are not independent
of one another and it is important to accommodate that in models of
the encounter process.  Let $x_{i}$ be the size of observation $i$ and
suppose $x_{i} \sim \mbox{Poisson}(\lambda)$. Then, the number of
groups of size $s$ is $\sum_{i} I\{x_{i} \equiv s\} = N_{s}$ and,
conditional on the total $N_{T} = \sum_{s} N_{s}$, the vector
$N_{1},\ldots,N_{S}$ has a multinomial distribution with sample size
$N_{T}$ and probabilities $\pi_{s} = \lambda_{s}/\sum \lambda_{s}$ as
before.  Thus, we can apply data augmentation to this case
directly.  Naturally, the encounter rate of groups should depend on
the size of the group which can be modeled in a number of ways, e.g.,
we can define
\[
\mbox{logit}(p_{s}) = \alpha_{0} + \alpha_{1} (s-1)
\]
or similar. 
As another example, consider the case of  a single sex-stratified
capture-recapture model. In this case, $S=2$, and the partially latent
variable $g$ equates to sex, and has two possible values (male/female). Such models have 
been fitted previously by data augmentation (Gardner et al. 2010; Mollet et
al. 2012).

\section*{Literature Cited}

\newcommand{\rf}{\vskip .1in\par\sloppy\hangindent=1pc\hangafter=1
                  \noindent}

\rf Alho, J.M. 1990. Logistic regression in capture-recapture models. 
{\it Biometrics} {\bf 46}:623-635.


\rf Borchers, D.L. and M.G. Efford. 2008.
Spatially explicit maximum likelihood methods for capture-recapture studies.
{\it Biometrics} {\bf 64}:377-385.

\rf Brooks, S.P. and A. Gelman. 1997. General methods for monitoring
convergence of iterative simulations. {\it Journal of Computational
  and Graphical Statistics} {\bf 7}:434-455. 

\rf Buckland, S., Anderson, D., Burnham, K., Laake, J., Borchers,
D. \& Thomas, L. 2001. Introduction to distance sampling: estimating
abundance of biological populations. Oxford University Press.

\rf  Chandler, R.B., J.A. Royle and D.I. King. 2011.
Inference about density and temporary emigration in unmarked populations.
{\it Ecology} {\bf 92}:1429-1435.

\rf Converse, S. J., G. C. White, and W. M. Block. 2006. Small mammal
responses to thinning and wildfire in ponderosa pine-dominated forests
of the southwestern United States. {\it Journal of Wildlife
  Management} {\bf 70}:1711-1722.

\rf Converse, S.J. and J.A. Royle. 2012. 
Dealing with incomplete and variable detectability in multi-year, 
multi-site monitoring of ecological populations. 
In "Design and Analysis of Long-term Ecological Monitoring Studies."

\rf Dorazio, R.M. and J.A. Royle. 2003.
Mixture models for estimating the size of a closed population when capture rates vary among individuals.
{\it Biometrics} {\bf 59}:351-364.

\rf  Dorazio, R.M., H.L. Jelks and F. Jordan. 2005.
Improving Removal-Based Estimates of Abundance by Sampling a Population of Spatially Distinct Subpopulations.
{\it Biometrics} {\bf 61}:1093-1101.

\rf Durban, J.W. and D.A. Elston. 2005.  Mark-recapture with occasion
and individual effects: abundance estimation through Bayesian model
selection in a fixed dimensional parameter space.  {\it Journal of
Agricultural, Biological, and Environmental Statistics} {\bf 10}:291-305.

\rf Fiske, I., and R.B. Chandler, R. 2011. \mbox{\tt unmarked}: An R
package for fitting hierarchical models of wildlife occurrence and
abundance. {\it Journal of Statistical Software} {\bf 43}:1-23.

\rf Gardner, B., J.A. Royle, M.T. Wegan, R.E. Rainbolt and P.D. Curtis. 2010.
Estimating black bear density using DNA data from hair snares.
{\it The Journal of Wildlife Management} {\bf 74}:318-325.

\rf Gelman, A and D.B. Rubin. 1992.  Inference from iterative simulation
using multiple sequences. {\it Statistical Science} {\bf 7}:457-511. 

\rf Gelman, A., Meng, X.L. and Stern, H. 1996.
Posterior predictive assessment of model fitness via realized discrepancies.
{\it Statistica Sinica} {\bf 6}:733-759.

\rf Gilks, W., A. Thomas, and D. Spiegelhalter. 1994.
 A language and program for complex Bayesian modelling.
{\it The Statistician} {\bf 43}:169--177.


\rf Huggins, R.M. 1989. On the statistical analysis of capture experiments.
{\it Biometrika} {\bf 76}:133-140.

\rf K\'{e}ry, M. and M. Schaub. 2011. 
Bayesian population analysis using WinBUGS: a hierarchical perspective.
Academic Press.

\rf Link, W.A. 1999. 
Modeling pattern in collections of parameters.
{\it The Journal of Wildlife Management} {\bf 63}:1017-1027.

\rf Link, W. A. and J. R. Sauer. 1997. 
Estimation of population trajectories from count data. {\it
  Biometrics} {\bf 53}:488-497.

\rf Link, W.A. and R.J. Barker. 2009. Bayesian 
Inference: With Ecological Applications.  Academic Press.

\rf Liu, J.S. and Y.N. Wu. 1999.
Parameter expansion for data augmentation.
{\it Journal of the American Statistical Association}
{\bf 94}:1264-1274.

\rf Mollet, P., M. K\'{e}ry, B. Gardner, G. Pasinelli and J.A. Royle. 2012.
Population size estimation for capercaille (Tetrao urogallus L.) using
DNA-based individual recognition and spatial capture-recapture
models. (in review).

\rf Pledger, S. 2000.
Unified maximum likelihood estimates for closed capture--recapture models using mixtures.
{\it Biometrics} {\bf 56}:434-442.

\rf Plummer, M. 2009.
JAGS: Just Another Gibbs Sampler.

\rf Royle, J.A., D.K. Dawson and S. Bates. 2004.
Modeling abundance effects in distance sampling.
{\it Ecology} {\bf 85}:1591-1597.

\rf  Royle, J. A. 2004a. N-mixture models for estimating population
size from spatially replicated counts.
{\it Biometrics} {\bf 60}:108-115.

\rf Royle, J.A. 2004b. 
Generalized estimators of avian abundance from count survey data.
{\it Animal Biodiversity and Conservation}
{\bf 27}:375-386.

\rf Royle, J.A. 2008. Hierarchical modeling of cluster size in wildlife surveys.
{\it Journal of Agricultural, Biological, and Environmental Statistics}
{\bf 13}:23-36.

\rf Royle, J.A. 2009.
Analysis of capture--recapture models with individual covariates using data augmentation.
{\it Biometrics}
{\bf 65}:267-274.

\rf Royle, J.A. and R.M. Dorazio. 2006.
Hierarchical models of animal abundance and occurrence.
{\it Journal of Agricultural, Biological, and Environmental Statistics}
{\bf 11}:249-263.

\rf Royle, J.A. and M. K{\'e}ry, M. 2007.
A Bayesian state-space formulation of dynamic occupancy models.
{\it Ecology} {\bf 88}:1813-1823.

\rf Royle, J.A., R.M. Dorazio and W.A. Link. 2007a. Analysis of multinomial
models with unknown index using data augmentation.
{\it Journal of Computational and Graphical Statistics} {\bf 16}:67-85.

\rf Royle, J.A., M. K{\'e}ry, M., R. Gautier, and H. Schmid. 2007b.
Hierarchical spatial models of abundance and occurrence from imperfect survey data.
{\it Ecological Monographs} {\bf 77}:465-481.

\rf Royle, J.A. and R.M. Dorazio. 2008.
Hierarchical Modeling and Inference in Ecology: The Analysis of Data
from Populations, Metapopulations and Communities.
Academic Press, San Diego, California, USA.

\rf Royle, J.A. and R.M. Dorazio. 2012.
Parameter-expanded data augmentation for Bayesian analysis of capture--recapture models.
{\it Journal of Ornithology} (Supp. 2) 1-17.

\rf  Royle, J.A. and K.V. Young. 2008.
A hierarchical model for spatial capture-recapture data.
{\it Ecology} {\bf 89}:2281-2289.

\rf Royle, J.A. 2009.
Analysis of capture - recapture models with individual covariates
  using data augmentation.
\emph{Biometrics} {\bf 65}:267-274.

\rf Schofield, M.R. and Barker, R.J. 2008.
A unified capture-recapture framework.
{\it Journal of Agricultural, Biological, and Environmental Statistics}
{\bf 13}:458-477.

\rf  Schofield, M.R. and Barker, R.J. 2011.
Full Open Population Capture--Recapture Models With Individual Covariates.
{\it Journal of Agricultural, Biological, and Environmental Statistics} {\bf
  16}:253-268.

\rf Schwarz, C. and A.~Arnason, 1996.
A general methodology for the analysis of capture-recapture
  experiments in open populations.
{\it Biometrics} {\bf 52}:860-873.

\rf Takemura, A. 1999.
Some superpopulation models for estimating the number of population uniques.
Proceedings of the International Conference on Statistical Data Protection SDP.
{\bf 98}:45-58.

\rf Tanner, M.A. and W.H. Wong. 1987.
The calculation of posterior distributions by data augmentation.
{\it Journal of the American Statistical Association}
{\bf 82}:528-540.

\rf Williams, B.K., J.D. Nichols, and M.J. Conroy, 2002.
Analysis and management of animal populations.
Academic Press, San Diego, California, USA.

\rf Yang, H.C. and A. Chao. 2005.
Modeling animals' behavioral response by Markov chain models for capture--recapture experiments.
{\it Biometrics} {\bf 61}:1010-1017.

\end{document}